\documentclass[%
 reprint,showkeys,
 amsmath,amssymb,
 aps,
]{revtex4-2}
\usepackage{graphicx}
\usepackage{dcolumn}
\usepackage{bm}
\usepackage{braket} 
\usepackage{hyperref}

\usepackage{subfigure}

\usepackage{overpic}
\usepackage{graphicx}
\usepackage{subcaption}
\usepackage{xcolor}
\usepackage{orcidlink}

\newcommand{\ketbra}[2]{\ensuremath{|{#1}\rangle\!\langle{#2}|}}

\newcommand{\tr}{\textnormal{Tr}}

\begin{document}

\preprint{APS/123-QED}

\title{Approaching the Limit of Quantum Clock Precision}

\author{Chad Nelmes$^{1,2,*}$\orcidlink{0009-0002-1686-6282}, Emanuel Schwarzhans$^{3,4}$\orcidlink{0000-0001-8259-9720}, Tony Apollaro$^3$\orcidlink{0000-0002-9324-9336}, Timothy Spiller$^{1,2}$\orcidlink{0000-0003-1083-2604}, and Irene D'Amico$^{1,2}$\orcidlink{0000-0002-4794-1348}}

\affiliation{$^1$School of Physics, Engineering and Technology, University of York, York, YO10 5DD, United Kingdom}

\affiliation{$^2$York Center for Quantum Technologies, University of York, York, YO10 5DD, United Kingdom}

\affiliation{$^3$Department of Physics, University of Malta, Msida MSD 2080, Malta}

\affiliation{$^4$ Atominstitut, Technische Universit{\"a}t Wien, 1020 Vienna, Austria}

\affiliation{$^*$Author to whom any correspondence should be addressed.}\thanks{Contact author: c.nelmes@york.ac.uk}
\date{\today}

\begin{abstract}
Precise and autonomous clocks are of fundamental interest and central importance to both foundational studies and practical applications. Here, we construct a blueprint for a quantum clock governed by time-independent interactions. By carefully-engineered
coherent transport in dissipative spin chains, we achieve a scaling exponent at the precision–resolution trade-off fundamental bound, bringing this within reach of physically realistic and experimentally accessible systems. We further introduce a sudden-quench protocol that enables repeated operation through a simple initialization and  detachment mechanism. Remarkably, the protocol is robust to imprecise detachment timing, implying that high-precision timekeeping can be achieved even when driven by a clock with much lower precision.
\end{abstract}

\maketitle

Throughout human history, precise timekeeping has been a central scientific and technical objective~\cite{orzel2022brief}. Atomic clocks \cite{Essen_1955} currently reach such extraordinary levels of precision that, if one had been operating since the first moments of the Big Bang, it would have accumulated an error of only about $1$ second today \cite{Aeppli_2024}. Such performance naturally raises a fundamental question: what ultimately limits the performance of a physical clock? The task of clocks is to provide a temporal reference. To achieve this, they must generate a signal without requiring any external control or timing, since this would presuppose prior knowledge about time and thus access to another clock~\cite{Milburn02042020,Woods2021,Erker2017,schwarzhans2020}. To achieve this, clocks necessarily have to feature a spontaneous, irreversible process. Assuming that the time signal comes in discrete packets called ``ticks", the above implies that the tick events are inherently stochastic and thus subject to fluctuations. 

In turn, the magnitude of these fluctuations limits the performance of clocks, in particular their stability over time, \emph{i.e.}, their precision (in the literature also referred to as accuracy  \cite{Manu}). Whilst fluctuations can be suppressed through careful engineering of the underlying quantum system~\cite{schwarzhans2020,Meier2025}, they cannot be eliminated: thermodynamic constraints akin to the thermodynamic uncertainty relations (TUR)~\cite{horowitz2020} impose a fundamental lower bound on tick fluctuations~\cite{Prech_2025}, as also demonstrated experimentally on a double quantum dot~\cite{Wadhia_2025}. Beyond this, any memory-less ``ticking clock" is subject to implementation-independent limitations, namely the fundamental precision–resolution trade-off (PRT), which bounds the maximally achievable precision of a clock for a given resolution (tick rate), where the precision scales inversely with the square of the resolution~\cite{Manu}.

Recent progress has clarified the precise mechanisms that allow us to improve clock performance, including the precise engineering of the systems' constituent couplings~\cite{Meier2025} and the utilization of temporal correlations in between ticks~\cite{Meier_2026}. However, to date, it is unclear how close one can realistically get to the fundamental PRT.
Here, we show that, by exploiting carefully-engineered coherent transport in a dissipative spin chain, the PRT bound can indeed be saturated: our model exhibits a precision which scales as the inverse square of the resolution. Remarkably, the proposed model is also physically implementable and experimentally feasible, entailing that the PRT is realistically achievable.

Spin chains are a natural platform for such transport dynamics, with a long history in quantum information processing as architectures for quantum state transfer and information transport~\cite{Wang_2001,Plenio_2005,Marta_2017,Bose,Christandl,Kay_2010,NikolopoulosJex2014,Hauke2016}. Recently, modulated 1-D spin chains with dissipation have also been proposed as autonomous quantum clocks~\cite{Meier2025,Meier_2026}. Here, we offer an alternative configuration, leveraging \textit{perfect state transfer} (PST) to optimize coherent transport, which brings us significantly closer to, and indeed allows us to reach, the theoretical upper-bound scaling of the PRT. 

Beyond the theoretical improvement upon previously achievable PRT curves, our approach remains experimentally accessible: it can reasonably be implemented in platforms capable of realizing the original set of tuned mirror-symmetric couplings for the implementation of PST, which have already been demonstrated in several physical platforms, from superconducting qubits to optical waveguides~\cite{Chapman_2016,Li2018,Xiang2024,Roy_2024,Wang_2025}. Our numerical simulations show the protocol performance for system sizes ranging from currently accessible chains of $\sim 10$ to $\sim 10^3$ qubits, providing a realistic outlook for both present and near-term experimental realizations.


\textit{Quantum state transfer and tick probability-} In this Letter, we use the PST protocol proposed in Refs.~\cite{Christandl,Niko} within the autonomous clock scenario: this way we achieve a PRT scaling equivalent to the theoretical upper bound. The PST protocol is implemented in one-dimensional, open-ended spin-$\frac{1}{2}$ chains modeled by the so-called XX-Hamiltonian
\begin{equation}
\hat{H}_{XX} =  \sum_{i=1}^{N-1} \frac{J_{i}}{2}\left(\hat{\sigma}_i^x\hat{\sigma}_{i+1}^x+\hat{\sigma}_i^y\hat{\sigma}_{i+1}^y\right).
    \label{Hami}
\end{equation}

Here, $N$ is the number of sites,  and $J_{i}$ describes couplings between adjacent qubits. The figure of merit for the efficiency of the state transfer, setting $\hbar=1$, is the fidelity
\begin{equation}
F(t) = \left|\langle N | e^{-i\hat{H}_{XX}t} | 1\rangle\right|^2, \quad 0 \leq F(t) \leq 1~,
\label{F}
\end{equation}
where the states are described in the single-particle basis $\ket{i}\equiv\ket{000\dots 1_i \dots000}$. By setting the inter-qubit couplings as
\begin{eqnarray}
J_{i} = J_0\sqrt{i(N-i)}, 
   \label{christa}
\end{eqnarray}
where $J_0$ sets the scale for the coupling amplitudes, PST, that is $F(t)=1$, is obtained at time $t_{PST} = \frac{\pi}{2J_0}$, which sets the speed limit for perfect end-to-end quantum communications within nearest-neighbor (NN) spin chains \cite{Yung_2006}.

We now wish to exploit the PST features of this chain to construct a clock approaching the PRT bound. To this aim, we couple the last site $N$ of the XX-model in Eq.~\ref{Hami} to a memoryless, dissipative environment at zero temperature. 
This open quantum system 
is effectively described by the Lindblad master equation
\begin{equation}
\dot\rho = -i[H_{\mathrm{XX}},\rho] + \Gamma( J\rho J^{\dagger} - \frac{1}{2}\{J^{\dagger}J,\rho\}),
\label{LME}
\end{equation}
where $J = \sqrt{\Gamma}\hat{\sigma}^-_N$ 
is the jump operator, and $\left\{\bullet,\bullet\right\}$ is the anticommutator.
The quantum jump unraveling of the master equation allows us to define an effective, non-Hermitian Hamiltonian governing the dynamics of the system between consecutive jumps
\begin{equation}
    \hat{H}_{\mathrm{eff}}
    = \hat{H}_{\mathrm{XX}}
    - \frac{i \Gamma}{2}\hat{\sigma}^+_N\hat{\sigma}^-_N,
    \label{EffH}
\end{equation}
where \( \Gamma > 0 \) is the decay rate associated with the sink \cite{Dalibard,Plenio_98,Manu,Meier2025}. 
Hence, as shown in Ref.~\cite{Meier2025}, the tick probability distribution function (PDF) is given by
\begin{align}
    \label{eq_tickpdf}
    p_{tick}(t)=\Gamma \left|\bra{N}e^{- i \hat{H}_{\mathrm{eff}}t}\ket{1}\right|^2~,~0\leq \int_{t_1}^{t_2}dt~p_{tick}(t) \leq 1
\end{align}
Although Eq.~\ref{F} and Eq.~\ref{eq_tickpdf} represent conceptually distinct quantities, their formal analogy is striking and we show in the following that the dynamics allowing PST can be leveraged for enhancing clocks' figures of merit. Exploiting the spectral properties of $\hat{H}_{XX}$, 
which result from the Hamiltonian 
being both symmetric ($\hat{H}_{XX}=\hat{H}_{XX}^{T}$) and persymmetric, and possessing chiral symmetry, we can recast the fidelity in Eq.~\ref{F} as
\begin{align}
&F(t)=\left|\sum_{\substack{k=-\frac{N}{2} \\ k\neq 0}}^{\frac{N}{2}} e^{-i \omega_k t} v_{k,N}v_{k,1}\right|^2=
4 \left|\sum_{k=1}^{\frac{N}{2}} (-1)^k v_{k,1}^2 \sin{ \omega_k t}\right|^2,
\label{F1}
\end{align}
where $\left\{\omega_k, \ket{v_k}\right\}$ are, respectively, the eigenvalues and eigenvectors of $\hat{H}_{XX}$.
From these equations we conclude that, in order to reach unit fidelity, the eigenvectors $\ket{v_k}$ need to have support both on the first and the last site, and their evolution should realize fully constructive interference at time $t_{PST}$, which is ensured by the linear spectrum $\omega_k$ yielding a dispersionless wave packet propagation along the chain. Similarly, the tick PDF in Eq.~\ref{eq_tickpdf} can be expanded in terms of the spectral decomposition of $\hat{H}_{\mathrm{eff}}$. The effective Hamiltonian in Eq.~\ref{EffH} is symmetric, which entails that the left eigenvectors are the transpose of the right eigenvectors, $\bra{l_k}=\left(\ket{r_k}\right)^T$, and possesses chiral symmetry. Hence, its complex eigenvalues satisfy $\epsilon_k=-\epsilon_{-k}^*$, and its right eigenvectors $\text{Re}\left[r_{kj}\right]+i \text{Im}\left[r_{kj}\right]=\left(-1\right)^{k+j-1}\left[r_{-kj}\right]+i \left(-1\right)^{k+j}\text{Im}\left[r_{-kj}\right]$. As a consequence, the tick PDF in Eq.~\ref{eq_tickpdf} can be cast as (see the End Matter)
\begin{align}
&\quad\quad\quad\quad\quad p_{tick}(t)=\Gamma \left|\sum_{\substack{k=-\frac{N}{2} \\ k\neq 0}}^{\frac{N}{2}} e^{-i \epsilon_k t} r_{k,N}r_{k,1} \right|^2 \nonumber\\
&= 4 \Gamma \left|\sum_{k=1}^{\frac{N}{2}} e^{\epsilon_k^I t} \left|r_{k,N}r_{k,1}\right|\sin\left(\epsilon_k^R t-\text{Arg}\left[r_{k,N}r_{k,1}\right] \right) \right|^2
\label{eq_tickpdf2}
\end{align}
where $\epsilon_k^R$ ($\epsilon_k^I$) is the real (imaginary) part of $\epsilon_k$. From Eq.~\ref{eq_tickpdf2}, we can set the analytic requirements for a clock based on our model to approach the PRT bound. That is, in order for the tick PDF to exhibit a pronounced peak (high precision) within a short time (high resolution), similar conditions as for PST apply: the right eigenvectors $\ket{r_k}$ need to have support both on the first and last site, and the real components of the eigenvalues must enable constructive interference quickly enough, before their imaginary components have suppressed the initial overlap $\left|r_{k,N}r_{k,1}\right|$. Clearly, the argument of the square modulus of Eq.~\ref{eq_tickpdf2} reduces to Eq.~\ref{F} for $\Gamma=0$ as $\hat{H}_{\text{eff}}=\hat{H}_{\text{XX}}$. 

\begin{figure}
    \centering
\includegraphics[width=\linewidth]{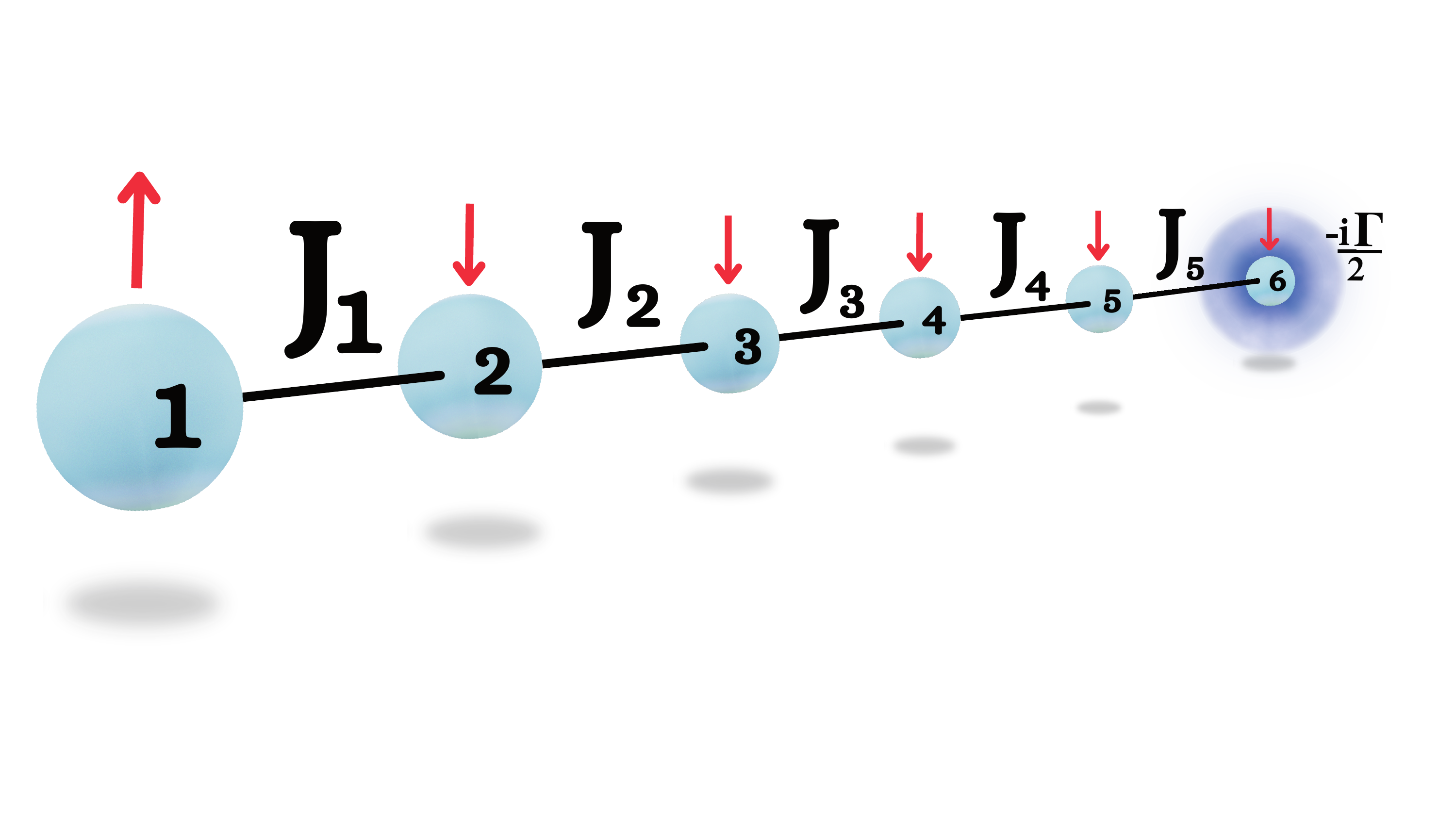}
    \caption{Clock set-up for an $N=6$-site chain. Red arrows indicate initial state $\ket{1}$, i.e. a localized excitation on the first site of a nearest-neighbor, time-independent Hamiltonian, with a sink (zero-temperature bath, blue sphere) on the final site. }
    \label{fig1}
\end{figure}

In the following we will set $\Gamma=1$ such that the sink rate is our characteristic energy and time scale, as well as for  consistency with prior works \cite{Meier2025,Meier_2026}. A schematic of the system is shown in Fig.~\ref{fig1}. Due to the non-Hermitian Hamiltonian, the state vector norm is not conserved, and therefore we instantiate a ``survival probability" given by
\begin{equation}
    S(t) = \ 
    \bra{1}{e^{i H_{\mathrm{eff}}^\dagger t} e^{-i H^{}_{\mathrm{eff}} t}}{\ket{1}}. 
    \label{surv}
\end{equation}
This survival probability (Eq.~\eqref{surv}) quantifies how much population remains in the ``no tick" subspace 
(i.e., not absorbed by the sink).

\textit{Clocks: Figures of Merit}- There are key aspects of a clock, whether it is classical or quantum, by which we may grade its performance. Here, we focus on the resolution $\nu = \mu^{-1}$, which is given by the inverse average time between ticks $\mu$, and the precision (or accuracy) $\mathcal{N}$, which quantifies the clock's stability over time. The survival probability $S(t)$ (Eq.~\eqref{surv}) plays the role of a cumulative non-tick
probability and induces a well-defined tick probability density via
$p_{\mathrm{tick}}(t) = -\dot S(t)$; hence, we may express its $n$-th moment as
\begin{equation}
\begin{aligned}
t_n &= n \int_0^\infty dt \; t^{n-1} S(t) \\
    &= (-1)^n n! \sum_{k,k'=1}^N 
       \frac{\langle 1\ketbra{l_k}{r_k} r_{k'}\rangle \braket{l_{k'}|1}}{[(\epsilon_k^{I}+\epsilon_{k'}^{I}) + i (\epsilon_k^{R}-\epsilon_{k'}^{R})]^n},
\end{aligned}
\label{moments}
\end{equation}
where $\bra{l_k}$ and $\ket{r_k}$ are the left and right eigenvectors of Eq.~\eqref{EffH}, such that $\braket{l_k|r_{k'}}=\delta_{kk'}$.

Using Eq.~\eqref{moments} we may calculate $\mu = t_1$, and the variance $\sigma^2=t_2-\mu^2$.
Detailed derivations of Eqs.~\eqref{moments}, $\mu$, and $t_2$ are provided in the End Matter. For a given clock architecture, we may now quantify the precision as 
\begin{equation}
     \mathcal{N} = \left(\frac{\mu}{\sigma}\right)^2.
     \label{precision}
\end{equation}
Assuming that the ticks are independently and identically distributed
(i.i.d.), we can interpret $\mathcal{N}$ as the average number of time units our clock ticks, until it is ``off" by exactly one tick. The region 
\begin{equation}
    \frac{\Gamma}{\nu}\leq \mathcal{N} \leq\frac{\Gamma^2}{\nu^2}
    \label{tradeoff}
\end{equation}
is still allowed by the PRT upper bound \cite{Erker2017,Manu,Meier2025}, but cannot be achieved by classically averaging a rate $\Gamma$ stochastic process, as averaging $M$ independent events yields a maximal precision of $\mathcal{N}=M$ and $\nu = \Gamma/M$ \cite{Manu}, yielding thus the LHS bound in Eq.~\eqref{tradeoff}. For a very good clock we require, with respect to the bound of Eq.~\eqref{tradeoff}, both high resolution and high precision.

\textit{Results-} Our findings emerge from a combined analytical–numerical approach, leveraging a well-known perfect-state-transfer solution (Eq.~\eqref{christa}) together with evolutionary computation. We use Differential Evolution (DE), a population-based stochastic optimization algorithm designed for continuous parameter spaces \cite{price2005differential}. 
It is particularly effective for cost functionals of highly non-convex functions, such as the survival probability in dissipative quantum spin chains. DE maintains a population of $M$ candidate solutions $\{\mathbf{x}_i\}_{i=1}^{M}$, each representing a vector of parameters. The algorithm is used to sustain and then minimize the survival probability, as defined in Eq.~\eqref{surv}, over a finite time window $T$ (in units of $\Gamma^{-1}$) for a given chain length $N$. We define the dynamical cost function 
\begin{equation}
    \mathcal{C}(\{J_{N-4}, \dots, J_{N-1}, J_0\}) = \sum_{t < T/2} (1 - S(t))^{2} + \lambda \sum_{t > T/2} S(t)^{2},
    \label{cost}
\end{equation}
where $\lambda > 0$ penalizes survival after the target tick time $T/2$, thus optimizing the precision of the clock within a finite window. The total simulation time window $T$ was chosen to scale as $\sim 10\sqrt{N}$, such that it safely encompasses the speed limit for NN chains \cite{Yung_2006}.

\begin{figure}[h!]
    \includegraphics[width=1.0\linewidth]{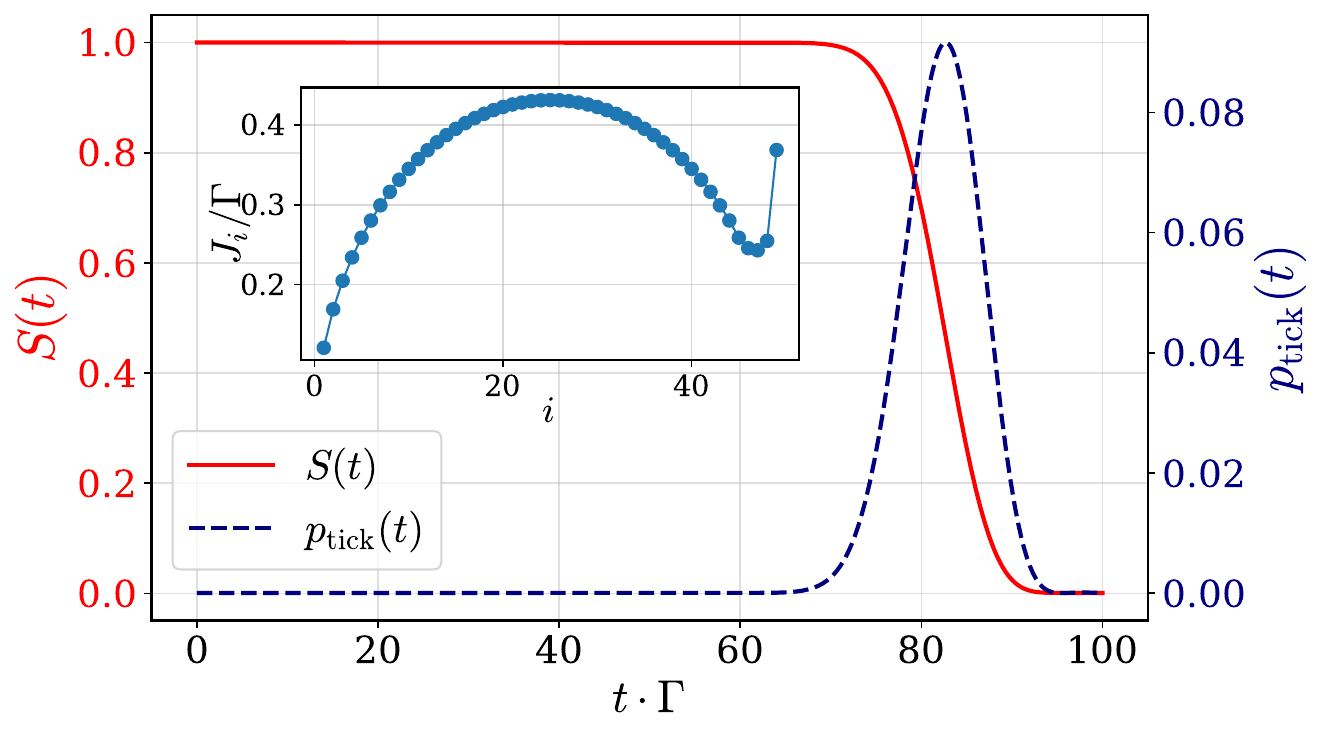}
            \caption{Survival probability $S(t)$ (red, left y-axis) and tick PDF $p_{\mathrm{tick}}(t)$ (blue dashed, right y-axis) for an $N=50$ chain with an all-but-four PST coupling (inset). Couplings are globally scaled with $J_0 = 0.0172$, and the final couplings up to the last site in the chain are $J_{[N-4:N-1]} \in [0.245,0.243,0.255,0.367]$. The inset shows the corresponding NN couplings $J_i/\Gamma$. This solution yields figures of merit of $\nu = 1.22\cdot10^{-2}$ and $\mathcal{N} = 361.62$}

 \label{fig2}
\end{figure}

Using a DE approach with the cost function  Eq.~\eqref{cost}, two parameters were optimized for each system size: the final four couplings up to the final site, and the global scaling factor $J_0$ defining the coupling profile in Eq.~(\ref{christa}). For a given $N$-site solution, as illustrated in Fig.~\ref{fig2}, we find that for the survival probability to decay to zero the final coupling $J_{N-1}$ must be enhanced relative to the preceding bond $J_{N-2}$. This effectively increases the rate at which the excitation is transferred from the penultimate site to the terminal site connected to the sink.
This finding is consistent with previous work \cite{Meier2025,Meier_2026}, which also presented chain-end apodization. 

\begin{figure}[h!]
    \centering
    \includegraphics[width=1.0\linewidth]{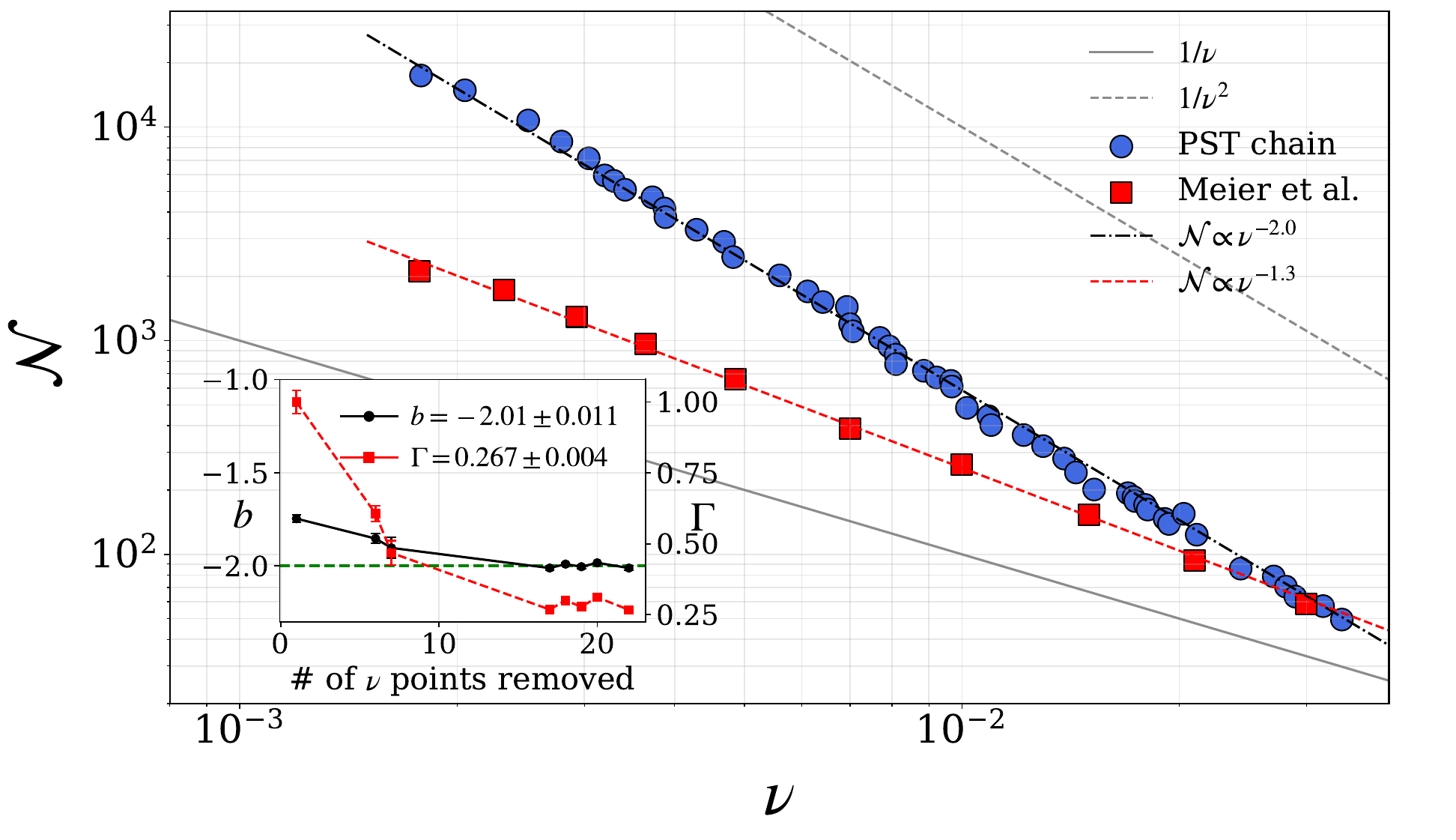}
    \caption{Log–log representation of the full optimization dataset (blue circles) spanning 
$N=10–2000$, with a direct comparison to the previous results (red squares) of \cite{Meier2025}, at commensurate system sizes. By removing finite-size effects (i.e. taking $N>10$) a scaling of $\mathcal{N}\propto\nu^{-2}$ emerges. Fits are for $N>10$. Convergency to -2 of the exponent $b$ in $N\propto\nu^b$ is shown in the inset, (left y-axis),  as lower$-N$ contributions are progressively removed. The right y-axis tracks the convergence of the intrinsic decay rate $\Gamma$.} 

    \label{fig3}
\end{figure}

The key result of this work is shown in Fig.~\ref{fig3}: we find a scaling in the precision of our all-but-four PST-coupled chains $\mathcal{N}\propto\nu^{-2}$:
this is equivalent in scaling to the PRT upper bound and significantly closer than the known results to-date for stochastic clocks with i.i.d. ticks \cite{Meier2025}.
Previous results yield a scaling of approximately $\mathcal{N} \propto \nu^{-\frac{4}{3}}$. 
This scaling avoids finite-size effects by considering large enough $N$ ($N>10$, see lower-left inset of Fig.~\ref{fig3}) which yield the overall trend of $\mathcal{N} \propto \nu^{-2}$. The couplings of the  numerically optimized chains exhibit a systematic scaling with the system size. Considering $N>10$, the global coupling scales as $J_0 \sim \,N^{-0.49}$, and the terminal coupling, which is always larger than the prior three optimized couplings, follows the scaling $\frac{J_{N-1}}{J_{\max}} \sim\,N^{-0.50}$, with $J_{\max}$ the chain largest coupling, indicating that its difference with respect to the adjacent coupling $J_{N-1}$ decreases with increasing chain size. Fig.7 in the End Matter shows the reported scaling.

\begin{figure}
    \centering
    \includegraphics[width=1\linewidth]{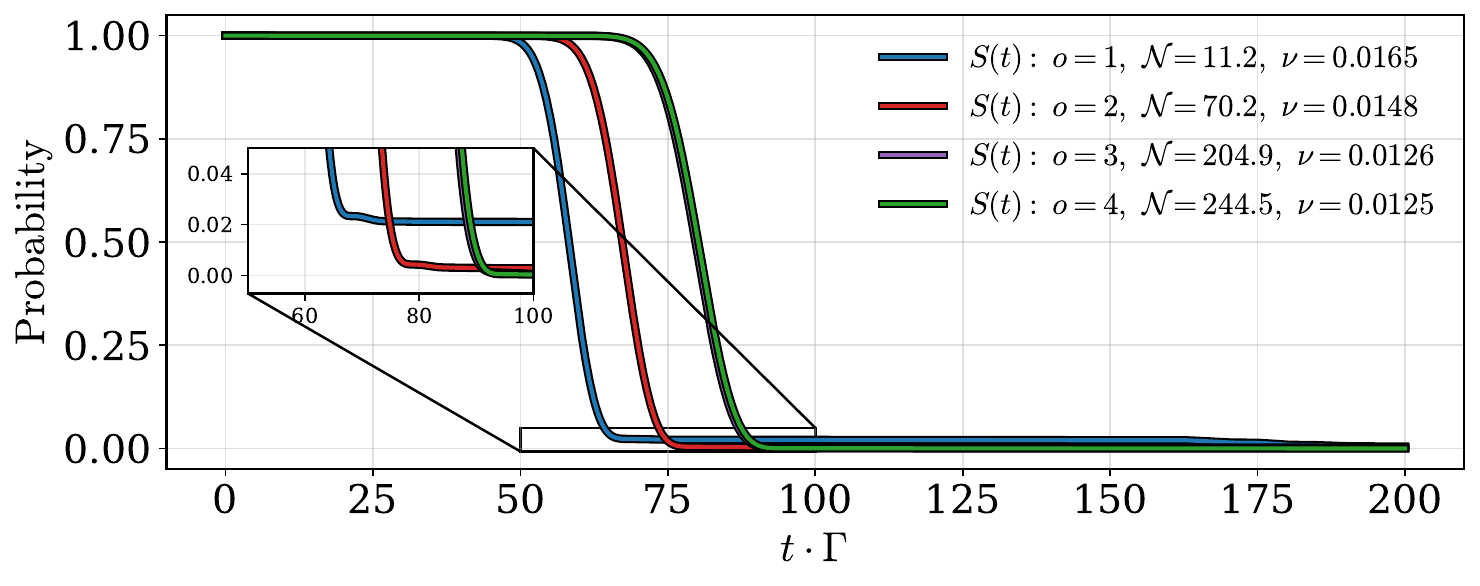}
    \caption{Survival probability dynamics trends  due to increased engineering via the number of couplings adjusted $o$, on the end of the chain for an optimized $N =40-$site XX chain with dissipation.
    }
    \label{fig_4}
\end{figure}

Coupling configurations, such as homogeneous or fully PST-coupled chains with a sink on the final site, yield significantly lower precision than the optimized couplings shown in Fig.~\ref{fig2}, motivating the need for the additional site-dependent optimization. We find that the number $o$ of optimized couplings required from the sink-site, while still enforcing the PST-coupling scheme throughout the rest of the chain, seems to reach optimality at four couplings, as the scaling of the PRT bound is achieved (Fig.~\ref{fig3}). By increasing the number of optimized couplings $o$ from 1 to 4, (see Fig.~\ref{fig_4}), the survival probability drops more decisively to zero and an increase in the precision for approximately the same resolution can be seen from three to four optimized couplings.

\begin{figure}[h!]
    \centering
    \includegraphics[width=1\linewidth]{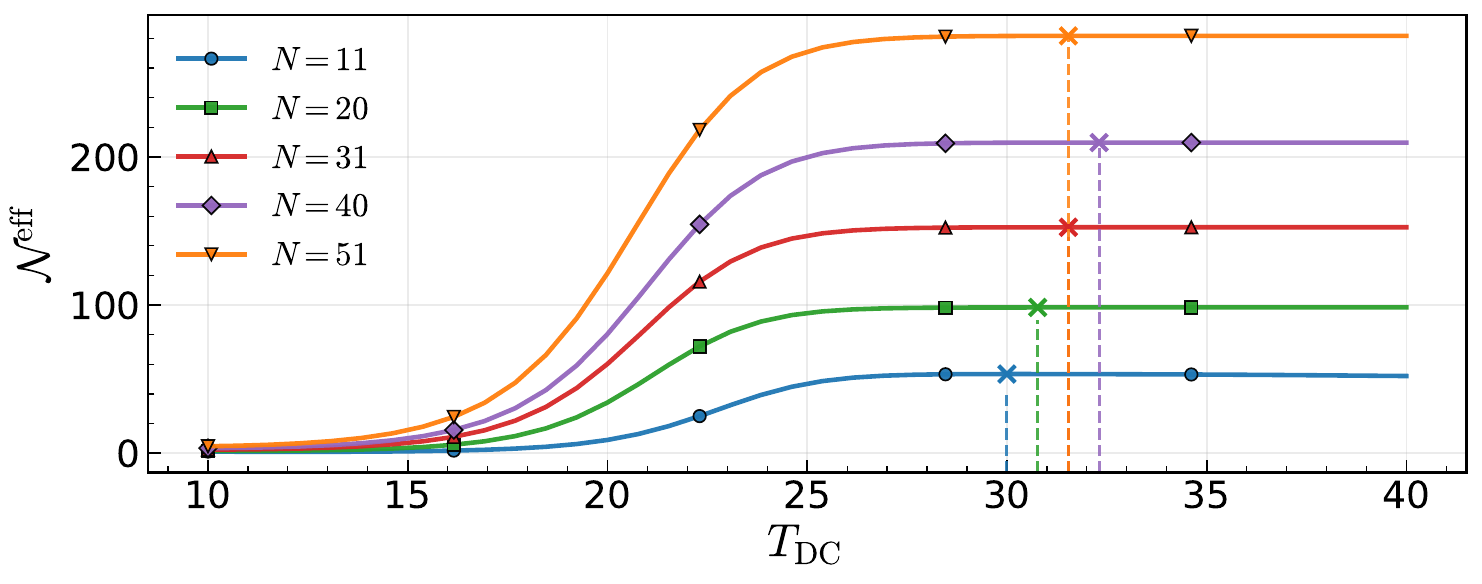}
    \caption{Effective clock precision as a function of decoupling time. 
The effective precision $N^{\mathrm{eff}}$ (Eq.~\eqref{precision}), extracted from the survival probability over the time at which the tick occurred for the optimized $N-$site solution, is shown as a function of the decoupling time $T_{\mathrm{DC}}$ in units of $\Gamma^{-1}$. Vertical dotted lines guide the eye to the points ($x$-markers) where the plateau of maximal accuracy attains 99\% of its original value, corresponding to particular even and odd $N$-solutions.
}
    \label{fig5}
\end{figure}

\textit{Experimental Feasibility and Clock Set-up }-Spin-chains with tunable coupling as described in the previous sections can be achieved via several quantum hardware, such as waveguide arrays through physical spacing \cite{Chapman_2016} or superconductors, where the effective inter-qubit coupling may be tuned by applying an alternating-current magnetic flux to modulate qubit frequencies \cite{Roy_2024}. Waveguides coupled to superconducting qubits may also effectively serve as a thermal bath, with their temperature modulated through the spectral density of microwave photons within the channel \cite{Yong_2022,AAmir_2025,sundelin2026quantum}.
While results of  previous sections combined with these hardware tunabilities may support the autonomous-clock ring-implementation-scheme proposed in \cite{Meier2025}, we now demonstrate a more-flexible protocol by which our system may be configured (and reconfigured) for continuous use. This protocol would allow to experimentally achieve the precision-vs-accuracy previously described, while departing from the autonomous-clock framework by using a ``low-precision clock" to support its practical experimental implementation.

Let us encode the first site with a ``spin-flip" and witness the natural dynamics generated by Eq.~\eqref{EffH} with the coupling configuration from Fig. \ref{fig2} with $o=4$: Fig.~\ref{fig5} shows that there is a {\it wide plateau} of decoupling times $T_{DC}$,  $T_{DC}\ll \tau_{2q}$,  when we may suddenly quench the Hamiltonian to decouple the first site from the rest of the chain without affecting the precision of the clock.
From data in Fig.~\ref{fig5} we find that the minimum $T_{\mathrm{DC}}$ approximately follows the advantageous scaling of $T_{\mathrm{DC}}/\mu\sim1/\sqrt{N}$. 
This property can be used to re-initialize-with-ease the clock, by  coarse control over the decoupling time and without affecting the absorption of the excitation within the sink. Experimentally, within superconducting qubit arrays, for example, the first site can be detached from the rest of the chain using dynamical decoupling techniques \cite{Li_2020,Tripathi_2022,Niu_2024}. The tick itself may be detected from the  waveguides via single-photon detection methods \cite{Kono_2018,anand_2025}, or through non-demolitionist continuous measurements of an eigenstate of the system \cite{Vijay_2011,Xin_2023}. Subsequent re-coupling of the first site to the rest of the chain can be triggered by positive detection of the tick. 

\textit{Conclusions and Outlook-} We have presented a linear spin chain protocol for timekeeping which exploits a modification to a well-known and experimentally achievable PST coupling profile. With a scaling exponent of -2, this configuration succeeds in achieving the PRT upper-bound scaling for i.i.d. ticks and presents a marked improvement to prior approaches. Furthermore, this scheme demonstrates robustness to sudden quenches of the system. Experimental platforms capable of realizing an effectively PST-coupled chain would provide a multifunctional resource, enabling both high-fidelity state transfer and precision timekeeping within a single architecture, provided the final qubit in the open-ended chain can also be coupled to a zero-temperature thermal bath. We have illustrated how this can be done within a superconducting platform coupled to waveguides.

Future work will explore alternative spin-network engineering schemes, hardware-specific energy landscapes, and varying topologies beyond NN couplings, with the aim of enabling precise timekeeping within more complex quantum hardware. Other schemes with very high state transfer fidelity \cite{Apollaro_2012,Alsulami_2022,Nelmes2025,Faria_2025} have been reported, both within spin chains and networks, which may also be investigated for the purposes of high-performance timekeeping. 
Interestingly, very recent clock based research \cite{Meier_2026} has found that if ticks are correlated, i.e., not i.i.d. as in our setup, then an exponential gain in the precision is found across ticks, so long as the chain is engineered appropriately. Expanding the results presented here, using a similar approach, will serve as the foundation of future work in support of the establishment of correlated quantum clocks.

\textit{Acknowledgments--}TJGA and IDA acknowledge funding from the Royal Society under the grant IES\textbackslash R3\textbackslash 243264 - International Exchanges 2024 Global Round 3. C.C. Nelmes acknowledges support from EPSRC, grant number is EP\textbackslash W524657\textbackslash1. ES and TJGA ackowledge funding from the European Union (Quantum Flagship project ASPECTS, Grant Agreement No. 101080167). Views and opinions expressed are however those of the authors only and do not necessarily reflect those of the European Union which cannot be held responsible for them.

\section*{End Matter}
\textit{Derivation of expressions for the moments of the tick probability distribution}-In the following we will derive the first and second moments of the tick probability density. These can be calculated analytically via the survival probability \cite{Manu} 
\begin{align}{S(t)=\tr[\ketbra{\psi(t)}{\psi(t)}]=\braket{\psi(t)|\psi(t)}}
\end{align}
as follows:
\begin{align}
    &\mu=\int_0^\infty S(t) dt\\&
    t_2=2\int_0^\infty t S(t) dt
\end{align}
The system is initialized in the state $\ket{\psi(0)}=\ket{1}$ in the site basis, meaning that there is one excitation at the first site only. It evolves according to the non-Hermitian effective Hamiltonian $H_\mathrm{eff}$ as
\begin{align}
    \ket{\psi(t)}=e^{-i H_\mathrm{eff} t}\ket{\psi(0)}
\end{align}
which can be decomposed as $H_\mathrm{eff}=\sum_k \epsilon_k |r_k\rangle\!\langle l_k|$, where $\bra{l_k}$ and $\ket{r_k}$ are the left and right eigenvectors, s.t. $\braket{l_k|r_k'}=\delta_{kk'}$ (note that in general $\ket{l_k}^\dagger \neq \bra{r_k}$ and $\braket{l_k|l_{k'}}\neq \delta_{kk'}$).
The survival probability can now explicitly be given as
\begin{align}
S(t)
  &= \operatorname{Tr}\!\bigl(e^{-i H_{\mathrm{eff}} t} |1\rangle\!\langle 1| e^{i H_{\mathrm{eff}}^\dagger t}\bigr)=\tr\bigl(e^{i H_{\mathrm{eff}}^\dagger t}e^{-i H_{\mathrm{eff}} t} |1\rangle\!\langle 1| \bigr)
  \\&=\sum_{i=1}^{N} \bra{i}e^{i H_{\mathrm{eff}}^\dagger t}e^{-i H_{\mathrm{eff}} t} |1\rangle\!\langle 1\!\ket{i}  \\&=\sum_{k,k'=1}^N e^{-i(\epsilon_k^R-\epsilon_{k'}^R)t} e^{-(\epsilon_k^I+\epsilon_{k'}^I)t} \,
     \langle 1|l_k\rangle \!\langle r_k | r_{k'} \rangle \!\langle l_{k'}|1\rangle
\end{align}
where the complex eigenvalues $\epsilon_k$ are decomposed into their real and imaginary parts $\epsilon_k^R$ and $\epsilon_k^I$ . Here, the first equality comes from the circularity of the trace. The last equality is due to the orthogonality of the position basis and substitution of the decomposed effective Hamiltonian. 
With this we can now calculate the first and second moments:
\begin{align}
\mu= \int_0^{\infty} S(t)\, dt
  = -\sum_{k, k^{\prime}=1}^N 
    \frac{\left\langle 1 \mid l_k\right\rangle
          \left\langle r_k \mid r_{k^{\prime}}\right\rangle
          \left\langle l_{k^{\prime}} \mid 1\right\rangle}
         {\left(\epsilon_k^I+\epsilon_{k^{\prime}}^I\right)+i\left(\epsilon_k^R-\epsilon_{k^{\prime}}^R\right)}
\end{align}
and
\begin{align}
t_2= 2\int_0^{\infty} t S(t)\, dt
  = 2\sum_{k, k^{\prime}=1}^N 
    \frac{\left\langle 1 \mid l_k\right\rangle
          \left\langle r_k \mid r_{k^{\prime}}\right\rangle
          \left\langle l_{k^{\prime}} \mid 1\right\rangle}
         {\left[\left(\epsilon_k^I+\epsilon_{k^{\prime}}^I\right)+i\left(\epsilon_k^R-\epsilon_{k^{\prime}}^R\right)\right]^2}
\end{align}
More generally, the $n$-th moment can be expressed as
\begin{align}
    t_n&=n\int_0^\infty dt \;t^{n-1}S(t)= \sum_{i=1}^N \bra{i} t^{n-1}e^{i H_\mathrm{eff}^\dagger t}e^{- iH_\mathrm{eff} t}\ketbra{1}{1}i\rangle \\
    &=n \sum_{k,k'=1}^N\int_0^{\infty} dt\; t^{n-1}e^{((\epsilon_k^{I}+\epsilon_{k'}^{I})+i (\epsilon_k^{R}-\epsilon_{k'}^{R}))t}\langle 1\ketbra{l_k}{r_k} r_{k'}\rangle \braket{l_{k'}|1}\\
    &=(-1)^n n!\sum_{k,k'=1}^N\dfrac{\langle 1\ketbra{l_k}{r_k} r_{k'}\rangle \braket{l_{k'}|1}}{[(\epsilon_k^{I}+\epsilon_{k'}^{I})+i (\epsilon_k^{R}-\epsilon_{k'}^{R})]^{n}}
\end{align}
To demonstrate that these are indeed real numbers, note that the following holds for conjugation:
\begin{align}
 \left\langle 1 \mid l_k\right\rangle^* &= \left\langle l_k \mid 1\right\rangle, \\
 \left\langle r_k \mid r_{k^{\prime}}\right\rangle^* &= \left\langle r_{k^{\prime}} \mid r_k\right\rangle, \\
 \left\langle l_{k^{\prime}} \mid 1\right\rangle^* &= \left\langle 1 \mid l_{k^{\prime}}\right\rangle.
\end{align}
For the denominator:
\begin{align}
\left[\left(\epsilon_k^I+\epsilon_{k^{\prime}}^I\right)+i\left(\epsilon_k^R-\epsilon_{k^{\prime}}^R\right)\right]^*
  &= \left(\epsilon_k^I+\epsilon_{k^{\prime}}^I\right)-i\left(\epsilon_k^R-\epsilon_{k^{\prime}}^R\right) .
\end{align}
Thus, it follows that
\begin{align}
\mu^*
  = \sum_{k, k^{\prime}=1}^N 
    \frac{\left\langle l_k \mid 1\right\rangle
          \left\langle r_{k^{\prime}} \mid r_k\right\rangle
          \left\langle 1 \mid l_{k^{\prime}}\right\rangle}
         {\left(\epsilon_k^I+\epsilon_{k^{\prime}}^I\right)-i\left(\epsilon_k^R-\epsilon_{k^{\prime}}^R\right)} .
\end{align}
Now we just swap the indices $k \leftrightarrow k^{\prime}$ in the sum for $\mu^*$. Since we sum over all $k, k'$, relabeling doesn not change the value:
\begin{align}
\mu^*
  = \sum_{k, k^{\prime}=1}^N 
    \frac{\left\langle l_{k^{\prime}} \mid 1\right\rangle
          \left\langle r_k \mid r_{k^{\prime}}\right\rangle
          \left\langle 1 \mid l_k\right\rangle}
         {\left(\epsilon_{k^{\prime}}^I+\epsilon_k^I\right)-i\left(\epsilon_{k^{\prime}}^R-\epsilon_k^R\right)} .
\end{align}
Now, noting that
\begin{align}
\left(\epsilon_{k^{\prime}}^I+\epsilon_k^I\right) &= \left(\epsilon_k^I+\epsilon_{k^{\prime}}^I\right), \\
-\left(\epsilon_{k^{\prime}}^R-\epsilon_k^R\right) &= \left(\epsilon_k^R-\epsilon_{k^{\prime}}^R\right),
\end{align}
and that the three scalar factors commute (as these are just complex numbers), so
\begin{align}
\left\langle l_{k^{\prime}} \mid 1\right\rangle
\left\langle r_k \mid r_{k^{\prime}}\right\rangle
\left\langle 1 \mid l_k\right\rangle
&=
\left\langle 1 \mid l_k\right\rangle
\left\langle r_k \mid r_{k^{\prime}}\right\rangle
\left\langle l_{k^{\prime}} \mid 1\right\rangle ,
\end{align}
this becomes
\begin{align}
\mu^*
  = \sum_{k, k^{\prime}=1}^N 
    \frac{\left\langle 1 \mid l_k\right\rangle
          \left\langle r_k \mid r_{k^{\prime}}\right\rangle
          \left\langle l_{k^{\prime}} \mid 1\right\rangle}
         {\left(\epsilon_k^I+\epsilon_{k^{\prime}}^I\right)+i\left(\epsilon_k^R-\epsilon_{k^{\prime}}^R\right)} .
\end{align}
This demonstrates that
\begin{align}
\mu^* = \mu
\end{align}
and thus $\mu$ is real. Similar analysis can be made for the second moment. 

\begin{figure}[h!]
    \centering
    \includegraphics[width=1\linewidth]{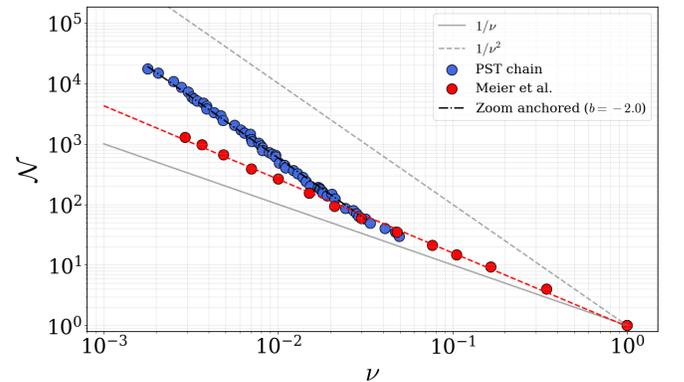}
    \caption{Full dataset within the bounds of the PRT presented in Fig.\ref{fig3}.}
    \label{fig_6}
\end{figure}

\textit{Overview of the Algorithm}-The DE algorithm evolves the population over $G$ generations using three main operations:

\begin{enumerate}

\item \textbf{Mutation:} For each target vector $\mathbf{x}_i$, three distinct population members $\mathbf{x}_a, \mathbf{x}_b, \mathbf{x}_c$ are randomly selected. A mutant vector $\mathbf{v}_i$ is constructed as
\begin{equation}
    \mathbf{v}_i = \mathbf{x}_a + F (\mathbf{x}_b - \mathbf{x}_c),
\end{equation}
where $F \in [0,2]$ is a mutation factor controlling the scale of perturbations.

\item \textbf{Crossover:} To increase diversity, a trial vector $\mathbf{u}_i$ is formed by combining the mutant vector $\mathbf{v}_i$ and the target vector $\mathbf{x}_i$ element-wise:
\begin{equation}
    u_{i,j} =
    \begin{cases}
        v_{i,j}, & \text{if } r_j < CR \\
        x_{i,j}, & \text{otherwise,}
    \end{cases}
\end{equation}
where $CR \in [0,1]$ is the crossover probability and $r_j$ are independent random numbers drawn uniformly from $[0,1]$.

\item \textbf{Selection:} The trial vector $\mathbf{u}_i$ replaces the target $\mathbf{x}_i$ if it yields a lower value of the cost function:
\begin{equation}
    \mathbf{x}_i^{\text{new}} =
    \begin{cases}
        \mathbf{u}_i, & \text{if } f(\mathbf{u}_i) < f(\mathbf{x}_i) \\
        \mathbf{x}_i, & \text{otherwise.}
    \end{cases}
\end{equation}
The best solution in the population is tracked at each generation.
\end{enumerate}
\textit{Log-log Plot of the Full Data within PRT-}Fig.~\ref{fig_6} shows the full dataset from  2-2000 sites for our chain described in the main text.

\begin{figure}[h!]
    \centering
    \includegraphics[width=1.0\linewidth]{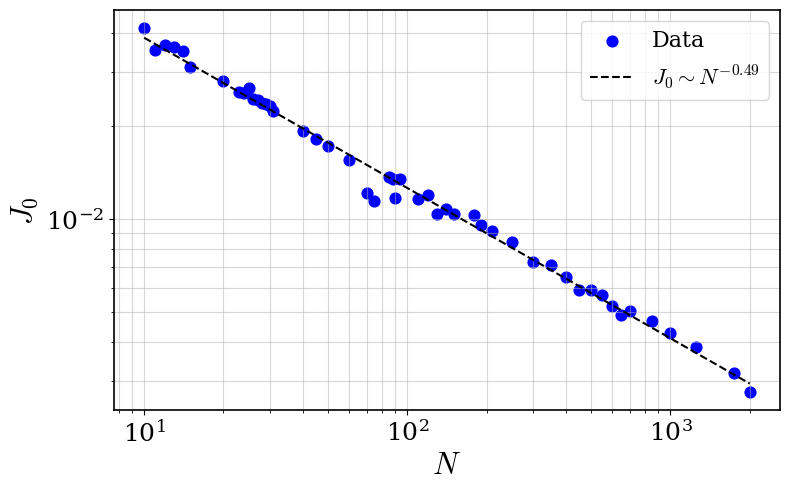}
    \includegraphics[width=1.0\linewidth]{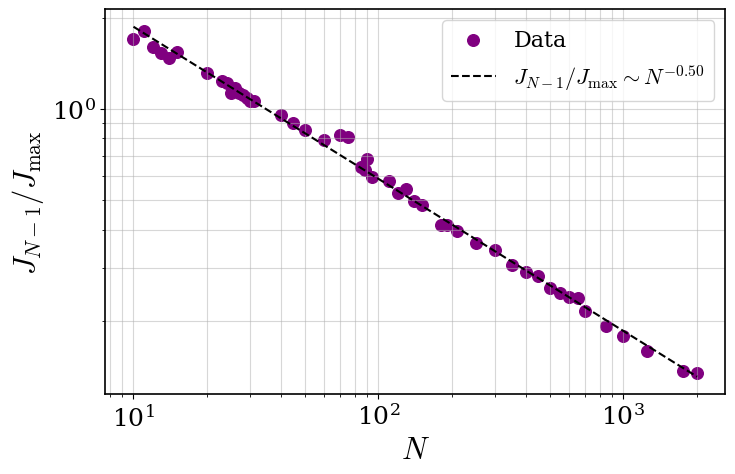}
    \caption{Log–log scaling of the bulk coupling $J_0$ with system size $N$. Ratio $J_{N-1}/J_{\rm max}$ illustrating the systematic enhancement of the boundary coupling required for efficient extraction as $N$ increases. The dotted line indicates fitting to the numerical data.}
    \label{fig:J_0scaling}
\end{figure}

\textit{Scaling of $J_0$ and $J_{N-1}$}-As reported in the main text, Fig.~\ref{fig:J_0scaling} reports the values of global scaling factor $J_0$ and the ration $\frac{J_{N-1}}{J_{max}}$ for varying lengths $N$ of the chain. Both values follow approximately an inverse-square-root scaling with $N$.

\textit{Derivation of the tick probability}-\label{app:ptickderivation}
The aim of this appendix is to show that the tick probability can be written as 

\begin{align}
& p_{\text {tick }}(t)=\Gamma\left|\sum_{k=-\frac{N}{2}}^{\frac{N}{2}} e^{-i \epsilon_k t} r_{k, N} r_{k, 1}\right|^2 \\
& =4 \Gamma\left|\sum_{k=1}^{\frac{N}{2}} e^{\epsilon_k^I t}| r_{k, N} r_{k, 1}|\sin \left(\epsilon_k^R t-\operatorname{Arg}\left[r_{k, N} r_{k, 1}\right]\right)\right|^2
\end{align}

The survival probability given by Eq.~\ref{surv} gives rise to tick PDF via 
\begin{align}
&p_\mathrm{tick}(t)= -\dfrac{d}{dt} S(t)= -\dfrac{d}{dt}\operatorname{Tr}[e^{-iH_\mathrm{eff} t }\rho (0) e^{-i H_\mathrm{eff}^\dagger t}]\\&= \operatorname{Tr}[i H_\mathrm{eff} e^{-iH_\mathrm{eff} t }\rho (0) e^{-i H_\mathrm{eff}^\dagger t}-e^{-iH_\mathrm{eff} t }\rho (0) iH_\mathrm{eff}^\dagger e^{-i H_\mathrm{eff}^\dagger t}]
\\&=-i\operatorname{Tr}[(H_\mathrm{eff} - H_\mathrm{eff}^\dagger)e^{-iH_\mathrm{eff} t }\rho (0) e^{-i H_\mathrm{eff}^\dagger t}].
\end{align}

Noting that 

\begin{align}
&H_\mathrm{eff}= H- \dfrac{i\Gamma}{2}\ketbra{n}{n}\implies
H_\mathrm{eff}-H_\mathrm{eff}^\dagger= - i\Gamma\ketbra{n}{n}
\end{align}
and that the initial state is $\rho(0)=\ketbra{1}{1}$ in the position basis, it follows that
\begin{align}
p_\mathrm{tick}(t)&= \Gamma \operatorname{Tr}[\ketbra{n}{n}e^{-i H_\mathrm{eff} t} \ketbra{1}{1}e^{i H_\mathrm{eff}^\dagger t}]
\\&= \Gamma \left| \bra{n} e^{-i H_\mathrm{eff} t} \ket{1}\right|^2.
\end{align}
We will set $\Gamma=1$ for simplicity. Decomposing the effective Hamiltonian as $H_\mathrm{eff}=\sum_{k=1}^N \epsilon_k\ketbra{r_k}{l_k}$ where $\ket{r_k}$ is and $\bra{l_k}$ are the right and left eigenvectors, fulfilling $\braket{l_k|r_q}=\delta_{kq}$ and assuming the eigenvalues are non-degenerate, we can write the tick PDF as
\begin{align}
&p_\mathrm{tick}(t)=\left|\sum_{k=1}^N e^{-i \epsilon_k t}\langle N\ketbra{r_k}{l_k}1\rangle\right|^2 
\\&=\left|\sum_{k=1}^N e^{-i \epsilon_k t}\langle N\ketbra{r_k}{r_k^*}1\rangle\right|^2 =\left|\sum_{k=1}^N e^{-i \epsilon_k t}r_{N,k}r_{1,k}\right|^2 .\label{eq:ptick_rnr1}
\end{align}
In the second equality we used the fact that $\ket{r_k}=\bra{l_k}^T$ (coming from $H_\mathrm{eff}=H_\mathrm{eff}^T$) and thus $\bra{l_k}\equiv\ket{l_k^*}^T=\bra{r_k^*}$ . Furthermore, $H_\mathrm{eff}$ is pseudo-Hermitian, meaning that its eigenvalues $\epsilon_k$ come in pairs
\begin{align}
 &\epsilon_k=\epsilon_k^R+i\epsilon_k^I\\&
  \epsilon_{-k}=-\epsilon_k^R+i\epsilon_k^I\; ,
\end{align}
where the we shifted the index by $N/2$, such that paired indices, $\{-k,k\}$, correspond to sign flips of the real component of their corresponding eigenvalues (assuming an even number for $N$). The superscripts $R$ and $I$ denote the real and imaginary part. This pairing of eigenvalues implies a corresponding pairing of eigenvector components, which we show now.

First, we define the operator
 \begin{align}
      \mathcal{T}=\sum_{n=1}^N(-1)^n|n\rangle\langle n| .
 \end{align}
 applied to $H_\mathrm{eff}$ it gives $\mathcal{T} H_\mathrm{eff}^* \mathcal{T}=-H_\mathrm{eff}$ due to the pseudo-hermiticity of the effective Hamiltonian.  
 Starting from the right-eigenvalue equation $ H_\mathrm{eff}|r_k\rangle = \epsilon_k|r_k\rangle$, take the complex conjugate: 
\begin{align}  H_\mathrm{eff}^*|r_k^*\rangle = \epsilon_k^*|r_k^*\rangle. \end{align}
Multiplying on the left by $\mathcal{T}$ and inserting $\mathcal{T}^2=\mathbf{1}$,
 \begin{align} (\mathcal{T} H_\mathrm{eff}^*\mathcal{T})(\mathcal{T}|r_k^*\rangle) = \epsilon_k^*(\mathcal{T}|r_k^*\rangle). \end{align}
Thus we get
\begin{align}  H_\mathrm{eff}(\mathcal{T}|r_k^*\rangle) = -\epsilon_k^*(\mathcal{T}|r_k^*\rangle) = \epsilon_{-k}(\mathcal{T}|r_k^*\rangle). \end{align}
Thus $\mathcal{T}|r_k^*\rangle$ is a right eigenvector with eigenvalue $\epsilon_{-k}$, and assuming non-degeneracy, which restricts its eigenvectors to lie on the same one dimensional eigenspace, it must be proportional to $|r_{-k}\rangle$:
\begin{align} |r_{-k}\rangle = s_k\,\mathcal{T}|r_k^*\rangle, \end{align}
 for some nonzero $s_k\in\mathbb{C}$. By using $\braket{r_{-k}^*|r_{-k}}=1$ it follows (from $\bra{r_{-k}^*}=(\ket{r_{-k}}^\dagger)^*$) that 
 \begin{align}\label{eq:s2}
 \bra{r_k}\mathcal{T} s_k^2 \mathcal{T} \ket{r_k^*}=s_k^2=1,
 \end{align}
 so $s_k=\pm 1$.  Taking the overlap with $\langle n|$ and using $\langle n|\mathcal{T} = (-1)^n\langle n|$, gives 
 \begin{align} r_{n,-k} = s_k(-1)^n r_{nk}^*. \end{align}

Setting $n=N$ and $n=1$ respectively gives the two identities needed below: 
\begin{align} r_{N,-k} = s_k(-1)^{N}\,r_{Nk}^*, \qquad r_{1,-k} = -s_k\,r_{1k}^*. \end{align}

Now we can reduce this back to positive indices. Using these identities and Eq.~\ref{eq:s2}, the product of components for the paired mode $-k$ is 
\begin{align}
&r_{N,-k}\,r_{1,-k} = -s_k^2(-1)^{N}\,r_{Nk}^*r_{1k}^* \\&= (-1)^{N-1}(r_{N,k}\,r_{1,k})^*. 
\end{align} 
For a chain of even length $N$ one has $(-1)^{N-1}=-1$, so writing $A_k := r_{N,k}\,r_{1,k}$, 

\begin{align}A_{-k}= r_{N,-k}\,r_{1,-k} = -A_k^*. \end{align} 

The contribution of the pair $\{k,-k\}$ to the amplitude is then
\begin{align} 
&e^{-i\epsilon_k t}A_k + e^{-i\epsilon_{-k}t}A_{-k}= e^{\epsilon_k^I t}\!\left(e^{-i\epsilon_k^R t}A_k - e^{+i\epsilon_k^R t}A_k^*\right) \\&
= -2i\,e^{\epsilon_k^I t}\,\mathrm{Im}\!\left(e^{-i\epsilon_k^R t}A_k\right). \end{align} Writing $A_k = |A_k|e^{i\,\mathrm{Arg}[A_k]}$, 
\begin{align}
&\mathrm{Im}\!\left(e^{-i\epsilon_k^R t}A_k\right) = |A_k|\sin\!\left(\mathrm{Arg}[A_k]-\epsilon_k^R t\right) \\&= -|A_k|\sin\!\left(\epsilon_k^R t - \mathrm{Arg}[A_k]\right), 
\end{align}
so each pair contributes $2i\,e^{\epsilon_k^I t}|A_k|\sin(\epsilon_k^R t - \mathrm{Arg}[A_k])$ to the amplitude. Summing over all $N$ which corresponds to summing over all $N/2$ positive-index pairs, taking the modulus squared as per Eq.~\eqref{eq:ptick_rnr1} and inserting $A_k$ gives  
\begin{align}
p_\mathrm{tick}(t) = 4\left|\sum_{k=1}^{N/2} e^{\epsilon_k^I t}|r_{N,k}\,r_{1,k}|\sin\!\left(\epsilon_k^R t - \mathrm{Arg}[r_{N,k}\,r_{1,k}]\right)\right|^2, 
\end{align}
which, restoring $\Gamma$, yields the desired result. $\blacksquare$

\end{document}